\documentclass[aps,prb,twocolumn]{revtex4-1}
\usepackage{amsfonts}
\usepackage{amsmath}
\usepackage{amssymb}
\usepackage[final,dvips]{graphicx}
\usepackage[sort&compress]{natbib}
\providecommand{\U}[1]{\protect\rule{.1in}{.1in}}
\begin{document}
\title{Linear magnetoconductivity in multiband spin-density-wave metals with nonideal nesting}
\author{A. E. Koshelev}
\affiliation{Materials Science Division, Argonne National Laboratory, Argonne, Illinois 60439}
\date{\today }

\begin{abstract}
In several parent iron-pnictide compounds the resistivity has an extended range of
linear magnetic field dependence. We argue that there is a simple and natural
explanation of this behavior. Spin density wave transition leads to Fermi-surface
reconstruction corresponding to strong modification of the electronic spectrum near
the nesting points. It is difficult for quasiparticles to pass through these points
during their orbital motion in magnetic field, because they must turn sharply. As
the area of the Fermi surface affected by the nesting points increases proportionally to magnetic field, this mechanism leads to the linear magnetoresistance. The crossover between
the quadratic and linear regimes takes place at the field scale set by the SDW gap
and scattering rate.
\end{abstract}
\maketitle

The studying of transport in magnetic field is the simplest way to characterize
electronic structure of new materials and quasiparticle scattering. The transport
properties of the recently discovered iron pnictides in magnetic field have some
anomalous features. In particular, the resistivity is found to have linear
dependence on magnetic field for several parent and underdoped compounds with
spin-density wave (SDW) long-range order, such as
CaFe$_{2}$As$_{2}$\cite{TorikachviliPhysRevB09}, BaFe$_{2}$As$_{2}$
\cite{IshidaPhysRevB11}, Ba(Fe$_{1-x}$Ru$_{x}$As)$_{2}$\cite{TanabePhysRevB11},
PrFeAsO \cite{BhoiAPL11}.

The linear magnetoresistance is not a new effect. It was first reported by Kapitza
for bismuth in 1928 \cite{KapitzaProc28}, see also \cite{YangSci99}, and later was
found in several other metals, see, e.g., Refs.\
\onlinecite{SimpsonJPhys73,XuNat97,BudkoPhysRevB98,LeePhysRevLett02,wangAPL12} and
discussion in the Pippard book \cite{PippardBook}. One can hardly expect a single
universal mechanism of this phenomenon. In different materials the linear
magnetoresistance may appear due to completely different reasons. In particular,
Abrikosov demonstrated that the linear dependence appears in the case of Dirac
electronic spectrum \cite{AbrikosovPhysRevB98,*AbrikosovPhysRevB99}. This mechanism
is frequently used for interpretation of the iron-pnictides data. Moreover, the
linear magnetoresistance sometimes is presented as a proof for the Dirac spectrum.

\begin{figure}[ptb]
\includegraphics[width=3.0in]{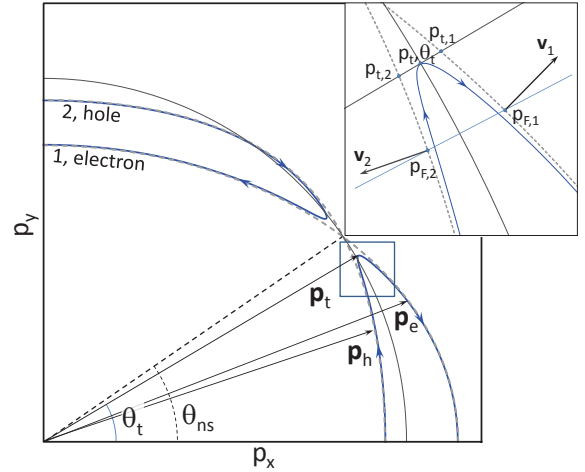}
\caption{Fermi surface in the region $p_{x},p_{y}>0$ for a two-band metal with the
SDW long-range order. Dashed lines show the bare Fermi surfaces. The electron Fermi
surface is displaced by the ordering wave vector $\mathbf{Q}$ to zone center. Arrows
show direction of the orbital motion in the magnetic field. The inset zooms into the
region near one turning point. }
\label{Fig-FermiSurfTurnPoint}
\end{figure}

We argue in this letter that the presence of the SDW order leads to a  simple and
natural mechanism for the linear magnetoresistance in the parent iron-pnictide
compounds, which, surprisingly, was not discussed. The SDW order mixes the electron
and hole bands which have different shapes. As a consequence, the nesting at the SDW
wave vector is only ideal at lines on the Fermi surface. Weak SDW order only
modifies electronic spectrum near these lines leading to reconstruction of the Fermi
surface, as illustrated in Fig.\ \ref{Fig-FermiSurfTurnPoint}. For fixed $p_{z}$
cross section the Fermi surface consists of four banana-shape pockets 
(only halves of two bananas are shown in Fig.\ \ref{Fig-FermiSurfTurnPoint}). Every pocket is
characterized by two sharp turning regions (banana tips) located near the nesting
points. In the magnetic field applied along $z$ direction the quasiparticles move
along the orbits located in the $xy$ plane. The turning regions where the orbits
transfer in between the electron and hole branches the velocity changes sharply and
smooth orbital motion is interrupted. This leads to enhanced quadratic
magnetoconductivity at small magnetic field and extended region of the linear
magnetoconductivity. The latter effect appears due to the linear growth with the field
of regular regions of Fermi surface affected by the turning regions. The crossover
field between the two regimes is proportional to the SDW gap $\Delta_m$ and inversely
proportional to the scattering time $\tau$. In the quadratic regime at small fields the turning-point contribution exceeds the conventional magnetoconductivity $\Delta \sigma (H)/\sigma(0) \sim (eH\tau/mc)^2$ by the factor $\sim \varepsilon_F/\Delta_m$, where $m$ is the effective mass and $\varepsilon_F$ is the Fermi energy.  This mechanism has been considered in Ref.\
\onlinecite{FentonSchofieldPRL05} for a metal with a single circular Fermi surface
reconstructed by commensurate density wave.

The  Fermi surface reconstruction caused by the magnetic transition in real
materials has been explored by
ARPES\cite{KondoPhysRevB10,*FuglsangJensenPhysRevB11,*YiNewJP12} and quantum oscillations\cite{TerashimaPhysRevLett11}. It may be rather
complicated. To illustrate the mechanism, we consider a simple two-band model with
the SDW order, see, e.g., Ref.\ \onlinecite{VorontsovPRB10}, which also has been used to describe the SDW transition in chromium \cite{RicePhysRevB70}. The model is
described by the Hamiltonian
\begin{equation}
\mathcal{H}=\mathcal{H}_{0}+\mathcal{H}_{AF}, \label{Hamilt}
\end{equation}
where the free-electron part is composed of the electron and hole contributions\footnote{In the electron part $\xi_{1,\mathbf{p}}$ the momentum $\mathbf{p}$ is measured with respect to the lattice wave vector $\mathbf{Q}$ at which the SDW ordering takes place.}
\begin{equation}
\mathcal{H}_{0}=\sum_{\mathbf{p},\sigma}\left(\xi_{1,\mathbf{p}}c_{\mathbf{p}\sigma}^{\dagger}c_{\mathbf{p}\sigma}+
\xi_{2,\mathbf{p}}d_{\mathbf{p}\sigma}^{\dagger}d_{\mathbf{p}\sigma}\right)
\label{HamFree}
\end{equation}
and the antiferromagnetic part is given by
\begin{equation}
\mathcal{H}_{AF}=-\sum_{\mathbf{p},\sigma}\sigma\Delta_{m}(c_{\mathbf{p}
\sigma}^{\dagger}d_{\mathbf{p}\sigma}+d_{\mathbf{p}\sigma}^{\dagger
}c_{\mathbf{p}\sigma}) \label{HamAF}
\end{equation}
with $\Delta_{m}$ being the SDW gap.

The simplest shapes of the free-electron spectra qualitatively describing iron
pnictides are parabolic bands,
\[
\xi_{1,\mathbf{p}}=\varepsilon_{1,0}-\mu+\frac{p_{x}^{2}}{2m_{x}
}+\frac{p_{y}^{2}}{2m_{y}},\ \xi_{2,\mathbf{p}}=\varepsilon_{2,0}-\mu
-\frac{p^{2}}{2m}.
\]
These bands are characterized by the Fermi momenta,
\[
p_{F,j}=\sqrt{2m_{j}\left(  \mu-\varepsilon_{1,0}\right)  },\
p_{F,h}\equiv p_{F,2}=\sqrt{2m\left(  \varepsilon_{2,0}-\mu\right)
}
\]
with $j=x,y$. The angular-dependent Fermi momentum for the electronic band is given
by $p_{F,1}(\theta)=\left(  \cos^{2}\theta/p_{F,x}^{2}+\sin^{2}
\theta/p_{F,y}^{2}\right)  ^{-1/2}$. In the further analysis, we assume that
for the selected $p_{z}$ cross section the inequality $p_{F,x}>p_{F,h}
>p_{F,y}$ holds. Introducing ratios, $r_{\alpha}=p_{F,h}/p_{F,\alpha}$ with
$r_{x}<1$ and $r_{y}>1$, we obtain that the ideal nesting is realized at the angles
satisfying
\[
\tan\theta_{\mathrm{ns}}=\sqrt{\frac{1-r_{x}^{2}}{r_{y}^{2}-1}}.
\]
These nesting angles depend on the z-axis momentum $p_z$ and trace the nesting lines
on the Fermi surface. To proceed, we will analyze the electronic spectrum near the
nesting angles in the presence of the SDW order.\footnote{A similar analysis has
been done in Refs.\ \onlinecite{BazaliyPhysRevB04,LinMillisPhysRevB05} to describe
the influence of the SDW order on the conductivity and Hall constant in chromium and
electron-doped cuprates.}

In the SDW state the quasiparticle spectrum has the following form
\cite{BazaliyPhysRevB04,LinMillisPhysRevB05,VorontsovPRB10}
\begin{equation}
E_{\mathbf{p},\pm}=\frac{\xi_{+}}{2}\pm\sqrt{\frac{\xi_{-}^{2}}{4}+\Delta
_{m}^{2}} \label{Spectrum}
\end{equation}
with $\xi_{\pm}=\xi_{1,\mathbf{p}}\pm\xi_{2,\mathbf{p}}$. We assume
$\Delta_{m}\ll\max_{FS}|\xi_{-}|$ so that the SDW order only modifies the spectrum
near the nesting angles. For the branches crossing the Fermi level the sign has to
be selected as $\pm\rightarrow\mathrm{sgn}(p_{F,1}-p_{F,2})$. The Fermi velocities
for the modified spectrum are
\begin{equation}
\mathbf{v}=\frac{\mathbf{v}_{+}}{2}\pm\frac{\mathbf{v}_{-}}{2}\frac{\xi_{-}
}{\sqrt{\xi_{-}^{2}+4\Delta_{m}^{2}}} \label{FermiVel}
\end{equation}
with $\mathbf{v}_{\pm}=\mathbf{v}_{1}\pm\mathbf{v}_{2}$ and $\mathbf{v}_{\alpha}=\partial \xi_{\alpha,\mathbf{p}}/\partial\mathbf{p}$. It is convenient to use the
polar coordinates $(p_{x},p_{y})=(p\cos\theta ,p\sin\theta)$ for fixed $p_{z}$ and
introduce the radial and angular components of the Fermi velocity,
$v_{\alpha,p}=\frac{d\xi_{\alpha,\mathbf{p}}}{dp}$,
$v_{\alpha,\theta}=\frac{1}{p}\frac{d\xi_{\alpha,\mathbf{p}}}{d\theta}$
where $\alpha$ is the band index. As the second band
is assumed to have the holelike spectrum, we have $v_{2,p}<0$.

In the case of weak SDW order, using linear expansion near the Fermi momenta $\xi_{\alpha,\mathbf{p}}\approx v_{\alpha,p}(p-p_{F,\alpha})$, we obtain
from $E_{\mathbf{p},\pm}\!=0$ the renormalized Fermi surface
\[
p_{F,\pm}\approx\frac{p_{F,1}\!+\!p_{F,2}}{2}\pm\sqrt{\frac{\left( p_{F,1}
\!-\!p_{F,2}\right)  ^{2}}{4}-\frac{\Delta_{m}^{2}}{v_{1,p}|v_{2,p}|}}
\]
(we omitted $\theta$ dependence in all $p_{F}$'s). It consists of four
banana-shaped sections, the two banana halves are shown in Fig.
\ref{Fig-FermiSurfTurnPoint}. Each section has two sharp turning points (tips
of banana). The angles of these turning points, $\theta_{t}$, can be found
from the following condition
\begin{equation}
\left\vert p_{F,1}(\theta_{t})-p_{F,2}(\theta_{t})\right\vert \approx\frac
{2\Delta_{m}}{\sqrt{v_{1,p}|v_{2,p}|}} \label{TurnAngle}
\end{equation}
and the Fermi momentum at the turning point is $p_{t}=\left(  p_{t,1}
+p_{t,2}\right)  /2$ with $p_{t,\alpha}=p_{F,\alpha}(\theta_{t})$. The Fermi surface
is eliminated in the angular range
\[
|\theta-\theta_{ns}|<|\theta_{t}-\theta_{ns}|\approx\frac{2\sqrt{v_{1,p}|v_{2,p}
|}\Delta_{m}}{\left[  \mathbf{v}_{1}\times\mathbf{v}_{2}\right]  _{z}p_{t}}.
\]
As the curvature of the Fermi surface sharply increases near the turning points, they have strong influence on transport properties at small magnetic fields.

Within the relaxation-time approximation for the Boltzmann equation, the
classical conductivity tensor for arbitrary magnetic field is given by
\begin{equation}
\sigma_{\alpha\beta}=2e^{2}\sum_{\mathrm{bands}}\int\frac{dp_{z}}{(2\pi)^{3}
}S_{\alpha\beta}(p_{z}), \label{CondGen}
\end{equation}
where $S_{\alpha\beta}(p_{z})$ describes the contribution from a single
$p_{z}$ slice
\begin{equation}
S_{\alpha\beta}\!=\!\frac{c}{eH}\int\frac{dp}{v}v_{\beta}\!\int_{p}
\frac{dp^{\prime}}{v^{\prime}}v_{\alpha}^{\prime}\exp\left(  \!-\!\int
_{p}^{p^{\prime}}\frac{dp^{\prime\prime}}{v^{\prime\prime}}\frac{c}{eH\tau
}\right) . \label{pzSlice}
\end{equation}
All $p$ integrals are performed along the fixed-$p_{z}$ orbits on the Fermi surface.
This presentation is similar to the so-called Shockley \textquotedblleft tube
integral\textquotedblright \cite{ShockleyPhysRev50,AbdelJawadNPhys06}. It goes
beyond the small-field expansion and provides a very convenient basis for analysis
of the conductivity especially when either the scattering rate $1/\tau$ or the Fermi
velocity $v$ have sharp features.

\begin{figure}[ptb]
\includegraphics[width=3.2in]{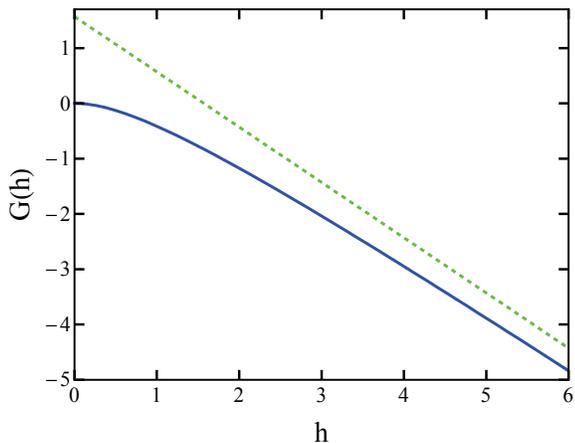}\caption{Function $G(h)$ in
Eq.\ (\ref{Gh}) which determines shape of magnetoconductivity. Dashed line
shows the high-field asymptotics.}
\label{Fig-Gh}
\end{figure}
We assume that the scattering rate is regular near the turning point and the anomalous behavior only appears due to modification of spectrum. This assumption definitely breaks down in the vicinity of the SDW transition point $T_{m}$ where scattering on the magnetic fluctuations becomes strong. The integral in the exponent of Eq.\ (\ref{pzSlice}) describes the orbital motion of the quasiparticles along the Fermi surface in the magnetic field. The SDW coupling forces the carriers to switch between the hole and electron orbits in the vicinity of the turning regions.

The simplest approximation is to treat the turning region as a point where the Fermi
surface has a sharp cusp and the velocity jumps. This approximation actually gives
correct result at high magnetic fields. We consider the contribution from the small
section of the Fermi surface near the turning point in which the orbital motion
starts at the point $\mathbf{p}_{h}$ of the hole branch and ends at the point
$\mathbf{p}_{e}$ of the electron branch (this sets the limits for the $p$ integral
in Eq. (\ref{pzSlice})). The distances from these points to
$\mathbf{p}_{t}=(p_{t,x},p_{t,y})$ are in the intermediate range: they are much
smaller than the Fermi momentum but the SDW corrections to the spectrum are assumed
already to be small. The former assumption allows us to neglect in this region the
bare curvature of the Fermi surface. In this case the direct calculation of the
contribution from one turning point gives $S_{xx}\! \approx\tau
v_{x,2}^{2}\frac{\left\vert \mathbf{p}_{t}-\mathbf{p} _{h}\right\vert }{v_{2}}+\tau
v_{x,1}^{2}\frac{\left\vert \mathbf{p} _{e}-\mathbf{p}_{t}\right\vert
}{v_{1}}+v_{x,2}\left( v_{x,1}\!-\!v_{x,2}\right) \frac{eH\tau^{2}}{c}$. For the
symmetric point $(-p_{t,x},p_{t,y})$ the orbital motion starts at the electronic
branch and ends at the hole branch. Therefore the contribution from this point
to the field dependent part is $S_{xx}(H)\!-\!S_{xx}(0)\!\approx \!v_{1,x}\left(
v_{2,x}\!-\!v_{1,x}\right)  eH\tau ^{2}/c$. Collecting contributions from the all
eight turning points, we obtain.
\begin{equation}
S_{xx}^{(\mathrm{tp})}(H)-S_{xx}(0)\approx-4\left( v_{2,x}-v_{1,x}\right) ^{2}
\frac{eH\tau^{2}}{c}
\label{LinSxxDirect}
\end{equation}
We see that treating the turning regions as sharp cusps leads to linear
magnetoconductivity.\footnote{Similar mechanism is discussed in the Pippard
book \cite{PippardBook}, p. 35, for square Fermi surface.}
As can be seen from Eq.\ (\ref{pzSlice}), this dependence appears because the Fermi momentum range within which quasiparticle can cross the turning point during its orbital motion is proportional to the magnetic field, $\Delta p=veH\tau/c$.

An accurate consideration should take into account a finite curvature of the Fermi
surface in the turning region. To perform $p$-integrals over the Fermi surface, we
have to find good parametrization. Due to the very simple dependence of velocity on
the parameter $\xi_{-}$, Eq.\ (\ref{FermiVel}), it is convenient to parametrize the
integration over the Fermi surface in terms of this parameter. For small shift of
$\mathbf{p}$ along the Fermi surface perpendicular to $z$ direction we have
$d\mathbf{p}=\frac{dp}{v}\mathbf{v} \times\mathbf{n}_{z}$. As
$d\xi_{-}=\mathbf{v}_{-}d\mathbf{p}$, using Eq.\ (\ref{FermiVel}) for velocity, we
straightforwardly derive the relation
\[
\frac{dp}{v}=\frac{d\xi_{-}}{\left[  \mathbf{v}_{1}\times\mathbf{v}
_{2}\right]  _{z}},
\]
which we can use to perform the integration over the Fermi surface in Eq.
(\ref{pzSlice}). In the vicinity of the nesting point we can neglect variations of
$\left[ \mathbf{v}_{1}\times\mathbf{v}_{2}\right]  _{z}$. Introducing the new
variable
\[
w=\frac{c}{eH\tau}\frac{\xi_{-}}{\left[  \mathbf{v}_{1}\times\mathbf{v}
_{2}\right]  _{z}}
\]
and the reduced field $ h=H/H_{\Delta}$ with the field scale
\begin{equation}
H_{\Delta}=\frac{2c\Delta_{m}}{e\tau\left[  \mathbf{v}_{1}\times\mathbf{v}
_{2}\right]  _{z}}, \label{MagFieldScale}
\end{equation}
we obtain $\xi_{-}/2\Delta_{m}=hw$ and
\[
\mathbf{v}(w)=\frac{\mathbf{v}_{+}}{2}\pm\frac{\mathbf{v}_{-}}{2}\frac{hw}
{\sqrt{h^{2}w^{2}+1}}.
\]
As a result, we obtain $S_{\alpha\beta}$ in Eq. (\ref{pzSlice}) in convenient
for calculation form
\[
S_{\alpha\beta}=\frac{eH\tau^{2}}{c}\int dwv_{\beta}(w)\int_{w}dw_{1}
v_{\alpha}(w_{1})\exp\left[  -\left(  w_{1}-w\right)  \right]  .
\]
We consider the $E_{\mathbf{p},+}$ branch located at $\theta<\theta_{t}$ which is
shown in the inset of Fig. \ref{Fig-FermiSurfTurnPoint}. For this branch the
velocity changes from the bare hole velocity $\mathbf{v} _{2,x}$ to the bare
electron velocity $\mathbf{v}_{1,x}$ as $w$ changes from large negative to large
positive values.

The contribution to the conductivity from one turning point in the $p_{z}$ slice,
\begin{align*}
&  S_{xx}(H)=\frac{2\tau\Delta_{m}h}{\left[  \mathbf{v}_{1}\times\mathbf{v}
_{2}\right]  _{z}}\int_{w_h}^{w_e} dw\int_{w}dw_{1}\exp\left[  -\left(  w_{1}
\!-\!w\right)  \right]  \\
&  \times\!\left(  \frac{v_{+,x}}{2}\!+\!\frac{v_{-,x}}{2}\frac{hw}{\sqrt
{h^{2}w^{2}\!+\!1}}\right)  \left(  \frac{v_{+,x}}{2}\!+\!\frac{v_{-,x}}{2}
\frac{hw_{1}}{\sqrt{h^{2}w_{1}^{2}\!+\!1}}\right).
\end{align*}
While the whole Fermi surface additively contributes to the zero-field conductivity,
the finite magnetoconductivity only appears due to the finite Fermi surface
curvature. As the turning regions have the largest curvature, they dominate in the
magnetoconductivity.  The total contribution from all eight turning points to the
field-induced change of $S_{xx}$ can be represented as
\begin{align}
&  S_{xx}^{(\mathrm{tp})}(H)-S_{xx}(0)=\frac{8v_{-,x}^{2}\tau\Delta_{m}
}{\left[  \mathbf{v}_{1}\times\mathbf{v}_{2}\right]  _{z}}G(h),\label{DSxxH}\\
&  G(h)=\int_{0}^{\infty}dx\int_{0}^{\infty}dy\exp\left(  -y\right)
\nonumber\\
&  \times\left(  \frac{x^{2}-h^{2}y^{2}/4}{\sqrt{\left(  x+h\frac{y}
{2}\right)  ^{2}\!+1}\sqrt{\left(  x-h\frac{y}{2}\right)  ^{2}\!+1}}
-\frac{x^{2}}{x^{2}\!+1}\right).  \label{Gh}
\end{align}
The dimensionless function $G(h)$ is plotted in Fig. \ref{Fig-Gh} and has the
following asymptotics
\[
G(h)=
\genfrac{\{}{.}{0pt}{}{-\frac{3\pi}{16}h^{2}\text{ for }h\ll1}{\frac{\pi}
{2}-h\text{ for }h\gg1}.
\]
The linear asympotics reproduces the result (\ref{LinSxxDirect}) obtained by direct
calculation. The typical field scale describing the crossover between these two
regimes is given by Eq.\ (\ref{MagFieldScale}).\cite{FentonSchofieldPRL05} It is
proportional to the SDW gap and inversely proportional to the scattering time.

Using presentation for the zero-field conductivity $\sigma_{xx}(0)=2e^{2}
\sum_{\mathrm{bands}}\int\frac{dp_{z}d\theta}{(2\pi)^{3}}m(\theta)v_{x}
^{2}(\theta)\tau(\theta)$, where $m(\theta)$ is the cyclotronic mass defined for the
bare bands, we can present the relative change of conductivity due to the turning points
at small fields,  $H\ll H_{\Delta}$, as
\begin{equation}
\frac{\sigma_{xx}^{(\mathrm{tp})}(H)-\sigma_{xx}(0)}{\sigma_{xx}(0)}=-\frac
{3}{32}\left(  \frac{eH\tau}{c}\right)  ^{2}\frac{
\left\langle v_{-,x}^{2}\left[ \mathbf{v}_{1}\times\mathbf{v}_{2}\right]_{z}\right\rangle }
{\left\langle mv_{x}^{2}
\right\rangle \Delta_{m}}. \label{RelSigHlow}
\end{equation}
Since the conventional magnetoconductivity can be estimated as $\left[
\sigma_{xx}^{(0)}(H)\!-\!\sigma_{xx}(0)\right]  /\sigma _{xx}(0)\!\approx\!-\left[
eH\tau/(mc)\right]  ^{2}$, we can see that the contribution from the turning points
exceeds the conventional one by the factor $\varepsilon_{F}/\Delta_{m}$. In the
linear regime, for $H\gg H_{\Delta}$, the relative change of conductivity can be
evaluated as
\begin{equation}
\frac{\sigma_{xx}^{(\mathrm{tp})}(H)\!-\!\sigma_{xx}(0)}{\sigma_{xx}(0)}
\!=\!
\frac{\left\langle \frac{v_{-,x}^{2}}
{\left[ \mathbf{v}_{1}\times\mathbf{v}_{2}\right]_{z}}\right\rangle}{\left\langle |m|v_{x}
^{2}\right\rangle }
\Delta_{m}  \!-\frac{\left\langle v_{-,x}^{2}\right\rangle }{\pi\left\langle |m|v_{x}
^{2}\right\rangle }\frac{\tau eH}{c}
\label{RelCondHhigh}
\end{equation}
We emphasize that the linear term does not depend on the SDW gap and coexists with the smaller quadratic contribution coming from the regular Fermi-surface. The linear behavior holds until $eH\tau/(mc)\ll 1.$

In conclusion, we considered magnetoconductivity for a multiband metal with the SDW
order. We demonstrated that, due to appearance of sharp turning points at the Fermi
surface, magnetoconductivity becomes linear when the magnetic field exceeds the
field scale proportional to the SDW gap and inversely proportional to the scattering
time. This mechanism provides a more natural explanation for the linear
magnetoresistance observed in iron pnictides
\cite{TorikachviliPhysRevB09,IshidaPhysRevB11,TanabePhysRevB11,BhoiAPL11} than the
popular mechanism based on Dirac spectrum. Taking typical values $v=10^7$cm/sec,
$\Delta_m=10$ meV, and $\tau=10^{-12}$ sec, we estimate $H_{\Delta}\approx 2$T, in
qualitative agreement with experiment. As both $\Delta_m$ and $\tau$ increase with
decreasing temperature, we can expect nonmonotonic temperature dependence of the
field scale. Namely, we can expect sharpening of the magnetic field dependence as
the temperature approaches the transition point due to decrease of $\Delta_m$ and at
low temperatures due to increase of $\tau$. Such behavior was not observed.
Typically, the field scale monotonically decreases with temperature
\cite{IshidaPhysRevB11,TanabePhysRevB11}, as expected when the temperature
dependence of $\tau $ dominates. However, no detailed study of magnetoresistance in
the close vicinity of the SDW transition was reported so far.

The localized Fermi surface reconstruction is not the only mechanism which can lead
to the linear magnetoconductivity. Alternatively, close to the transition point the
scattering caused by the antiferromagnetic fluctuations leads to suppression of the
relaxation time near the nesting points (``hot spots''). This also gives the linear
magnetoconductivity due to the interruption of the orbital motion
\cite{RoschPhysRevB00}. The scattering mechanism clearly becomes dominant as
the temperature approaches $T_{m}$. The crossover between the two mechanisms in the vicinity of the transition point is an interesting topic for future study.

I would like to thank Vivek Mishra for useful discussions. This work was supported
by UChicago Argonne, LLC, operator of Argonne National Laboratory, a U.S. Department
of Energy Office of Science laboratory, operated under contract No.
DE-AC02-06CH11357.
\bibliographystyle{apsrev4-1}
\bibliography{MagnCondSDW}

\end{document}